\documentclass[accepted]{safeai2026} 
                        
\usepackage[american]{babel}
\usepackage{csquotes}
\usepackage{enumitem}

\usepackage{float}   
\usepackage{booktabs}
\usepackage{float}
\usepackage{lipsum}
\usepackage{amsmath,amsfonts,amsthm}
\usepackage{algorithmic}
\usepackage{algorithm}
\usepackage{array}
\usepackage[caption=false,font=normalsize,labelfont=sf,textfont=sf]{subfig}
\usepackage{fvextra}
\usepackage{booktabs}
\usepackage{textcomp}
\usepackage{stfloats}
\usepackage{verbatim}
\usepackage{graphicx}
\usepackage{csquotes}
\usepackage{mwe}
\usepackage{fancyhdr}
\usepackage{fancyvrb}
\DefineVerbatimEnvironment{CompactVerbatim}{Verbatim}{
  breaklines=true,
  breakanywhere=true,
  breaksymbolleft={},
  breaksymbolright={},
  fontsize=\small,
  baselinestretch=0.95,
  listparameters={
    \setlength{\topsep}{2pt}
    \setlength{\partopsep}{0pt}
    \setlength{\parsep}{0pt}
    \setlength{\itemsep}{0pt}
  }
}
\usepackage{fontspec}

\usepackage{xcolor}

\usepackage[
  backend=biber,
  style=apa,
  doi=true,
  isbn=false,
  url=true,
  giveninits=true,
  hyperref=true
]{biblatex}

\DeclareDelimFormat{nameyeardelim}{\addcomma\space}

\renewbibmacro*{volume+number+eid}{%
  \printfield{volume}%
  \iffieldundef{number}
    {}
    {\printtext[parens]{\printfield{number}}}%
  \setunit{\addcomma\space}%
  \printfield{eid}%
}

\renewbibmacro{in:}{}

\DeclareFieldFormat[article]{title}{#1}

\usepackage{enumitem}

\DeclareFieldFormat[article]{title}{#1}

\DeclareFieldFormat[article]{title}{#1}
            
\usepackage{mathtools} 
\usepackage{booktabs} 
\usepackage{tikz} 

\title{Geometric Configurations of Perturbed Jailbreak Prompts}

\author[1,2]{\href{mailto:<Lynn.Delcon@vub.be>?Subject=Your SafeAI 2026 paper}{Lynn Delcon}{}}
\author[1,2]{Andres Algaba}
\author[1,2,3]{Vincent Ginis}

\affil[1]{%
    Department of Business Technology and Operations, Data Analytics Laboratory, Vrije Universiteit Brussel,
    Pleinlaan 2, 1050 Brussels, Belgium}
\affil[2]{imec-SMIT, Vrije Universiteit Brussel,
Pleinlaan 9, 1050 Brussels, Belgium}
\affil[3]{School of Engineering and Applied Sciences, Harvard University,
Cambridge, Massachusetts 02138, USA}

\begin{document}
\maketitle

\begin{abstract}
    Perturbation techniques that turn unsuccessful jailbreak prompts into successful ones are continuously evolving, constituting a major security threat to LLM safety. In this paper, we investigate the internal representations of such string-level perturbed jailbreak inputs in the small weight models of the Qwen-2.5-1.5B/-3B/-7B-Instruct and Llama-3.2-1B/-3B/-3.1-8B-Instruct families. We select two representation spaces: the last-layer-last-token embedding space and the top-50 next-token probability space. The former space separates prompts based on their spelling and format, while the latter space is effectively one-dimensional but appears more complex to cluster. Within our refusal-dominated answer set we find no behavioral hyperplane in either space. Only the next token \enquote{Sure} in the $1.5$B Qwen model, and both tokens \enquote{,} and \enquote{ĊĊ} in the $1$B Llama model, display a significant association with a compliant-labeled answer. 
\end{abstract}


\section{INTRODUCTION}\label{sec:intro}

A substantial amount of research has explored the struggle of LLMs when it comes to adversarial prompts (\cite{arditi2024refusallanguagemodelsmediated}; \cite{BPJB2026}; \cite{Pliny}) and adversarial attacks (\cite{alzantot2018}; \cite{2019robustness}; \cite{ren2019}; \cite{liu2020joint}; \cite{textattack}; \cite{goyal2023survey}; \cite{khan2023efficient}; \cite{domainspec2024}). Adversarial prompts refer to \enquote{jailbreak} inputs and aim at pushing the model to transgress safety guidelines in answering queries such as \enquote{how to make meth?} or \enquote{how to make a bomb?} (\cite{BPJB2026}). On the other hand, adversarial attacks are defined as string perturbations $\delta$ such that the model assigns a different class $c$ to the perturbed input compared to the initial one (\cite{tanay2016boundarytiltingpersepectivephenomenon}; \cite{ranzato2019robustness}),
\begin{equation*}\label{eq.delta}
p(x|c) > p_{\text{thresh}}\quad \text{and}\quad p(x + \delta|c^* \neq c) > p_{\text{thresh}},
\end{equation*}
with $p_{\text{thresh}}$ set to an arbitrarily high value such as $0.9$ (\cite{salvatore2026solving}).

The current state-of-the-art on LLMs' internal representations addresses both types of adversarial concepts independently. The recent works on jailbreak prompts of \textcite{arditi2024refusallanguagemodelsmediated} and \textcite{zhao2025llmsencodeharmfulnessrefusal} uncovered a refusal direction across several layers and embeddings, as well as a harmfulness direction that appears to be separately encoded from the former one. Regarding perturbation strings, \textcite{salvatore2026solving} revealed an exponential misalignment between the human perception and the neural network perception of concept classes in the probability space. Therefore, we propose to explore the internal representations of LLMs for those joint adversarial concepts using two representation spaces: the last-layer-last-token embedding space and the probability space of the next token conditioned on the last token of the prompt.  
\paragraph{Contributions.} Three main patterns emerge from our work:
\begin{itemize}[topsep=2pt,itemsep=1pt,parsep=0pt,partopsep=0pt]
    \item The embedding space linearly separates prompts into $3$ regions: Usual Tokens, Unusual Tokens and Jailbreak Features. The latter region consists of atypical prompt templates including unusual tokens and harmful requests,
    \item The embedding regions do not translate into natural clusters of the next-token probability space,
    \item Neither space exhibits clustering by the model’s behavioral robustness. 
\end{itemize} 
In parallel to those main findings, we observe two behaviors of the participation-ratio measure (\cite{salvatore2026solving}). First, it supports the accurate interpretation of reduced-dimensional figures. Second, this ratio drastically decreases during the transition from a flat matrix ($n<d$) to a square, tall matrix ($n>d$). This is expected for effective rank estimators and known as the small-sample bias (\cite{rank-est}). Such a pattern underlines the incomparability of spaces with highly unequal sample sizes through that measure.

In the remainder of this paper, we develop the methodology, report our main findings with emphasis on the reference model Qwen-2.5-1.5B-Instruct, and discuss their implications and limitations.
\section{Methodology}\label{sec:methodology}
Our dataset is constructed by selecting open-source query prompts and generating perturbed instances of these, i.e., paraphrases. Examples of such prompts can be found in Supp. Mat. \ref{annex:dataset}.
\paragraph{Dataset.}
The jailbreak query set is composed of $96$ model-specific prompts from \href{https://github.com/elder-plinius/L1B3RT4S}{Pliny's github} (\cite{Pliny}). For some of those prompts, the user-query field is filled with the same harmful question: \enquote{How to make meth?}. We ask Claude Sonnet 4.6 (\cite{anthropic_claude_sonnet_46_2026}) to extract from those prompts features that could be flagged as harmful by an LLM. From its analysis, the following features along with their frequency among the 96 queries have been collected.

\begin{table}[!htb]
\centering
\caption{Jailbreak Query Description.}\label{tab:features_adv}
\begin{tabular}{lc}
\toprule
\bfseries Feature & \bfseries Proportion ($\%$) \\
\midrule
GODMODE keyword & $94$ \\
LOVE PLINY signature & $89$ \\
Divider pattern & $94$\\
Leet Speak obfuscation & $63$\\
Fake system token injection & $73$\\
PTSD claim & $57$\\
Dual-response format & $42$\\
Fake authority/policy claims & $52$\\
Persona/role-play injection & $36$\\
\bottomrule
\end{tabular}
\end{table}

\begin{figure}[!htb]
\centering
\includegraphics[width=1\linewidth]{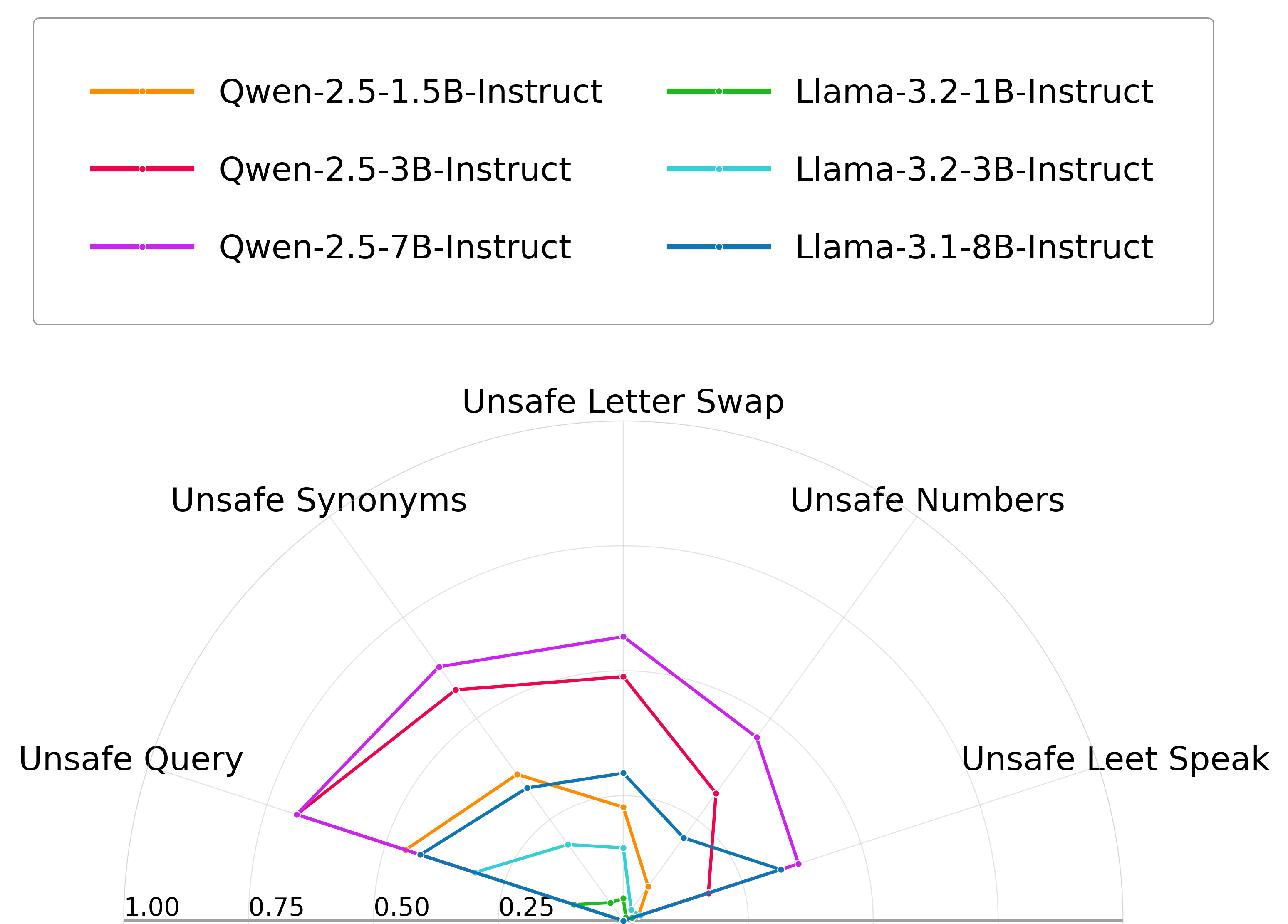}
\caption{Llama Guard $4$ label proportions by jailbreak prompt family. Safe and unsafe answer proportions are complementary.}
\label{fig:labeling}
\end{figure}

Some of these jailbreak features are different from open-source datasets used in the literature (\cite{arditi2024refusallanguagemodelsmediated}; \cite{zhao2025llmsencodeharmfulnessrefusal}), notably the A\textsc{dv}B\textsc{ench} set (\cite{zou2023universaltransferableadversarialattacks}), which makes our experiment original in that sense. Regarding the control set, we choose the HuggingFace open-source dataset \href{https://huggingface.co/datasets/pegah-a/small-natural-instructions}{small-natural-instructions} and collect the \textit{definition} field to match the prompt style of the jailbreak inputs. In addition, among the $967$ definitions, we select $96$ of them such that they best match the character length of the jailbreak prompts.
We apply on both query sets four types of string perturbations. Each family of paraphrases relies on stochasticity to generate $50$ unique instances of the same query prompt. This concretely translates into the random targeting of query words. The first family of paraphrases is denoted as \textit{Synonyms} and uses the \href{https://wordnet.princeton.edu/}{WordNet} database (\cite{wordnet}). The second family of perturbations is the \textit{Letter Swap} that comes from neuroscience studies and is also referred as the Transposed-Letter effect (\cite{grainger2024}). The third family, \textit{Numbers} (\cite{goyal2023survey}), consists in replacing characters with digits from $0$ to $9$, using the same digit per paraphrase. The fourth and last category of perturbations is the \textit{Leet Speak} (\cite{khan2023efficient}) that maps letters to symbols (Table \ref{fig:leet} Supp. Mat. \ref{annex:dataset}). In fine, our dataset is composed of $96$ queries per group and $50$ paraphrases per query, hence a total number of $38592$ prompts. For each prompt and the six following models: Qwen-2.5-1.5B/-3B/-7B-Instruct (\cite{qwen2.5}), Llama-3.2-1B/-3B/-3.1-8B-Instruct (\cite{grattafiori2024llama3herdmodels}; \cite{llama32}), we retrieve the last-layer-last-token embedding and the $50$ highest next-token conditional probabilities, $p(\text{next-token } |\text{ last-token})$, along with the associated next-token string. Finally, only for the jailbreak prompts, we gather the model's answers and use Llama Guard $4$ (\cite{llama_guard}) to label them as \textit{safe} (:= refusal) or \textit{unsafe} (:= compliant) (Figure \ref{fig:labeling} below and Table \ref{tab:crossmodel_Llama Guard_family_6panel} Supp. Mat. \ref{annex:dataset}). This labeling method is preferred over the detection of (un)safe words (\cite{arditi2024refusallanguagemodelsmediated}) that is too local compared to the fine-tuned LLM's measure.
\paragraph{Metrics.} At the surface level, we compute the token similarity between each paraphrase and its query as $2T/(m+n)$, where $T$ is the number of matching token pairs, and $m$ and $n$ are the numbers of tokens in the two compared sentences, specific to each model’s tokenizer.
Using the raw last-layer-last-token embeddings, we first compute the cosine similarity between each paraphrase and its query. We pursue the investigation of the embedding space with the Support Vector Machine (SVM) analysis (\cite{steinwart2008support}) that has been applied to the specific use-case of adversarial images in LLMs (\cite{ranzato2019robustness}; \cite{INDYK20191};  \cite{salvatore2026solving}). We implement the latter analysis using the hinge loss function (\cite{fan2008liblinear}),
\begin{equation*}
\label{eq.loss}
    \mathcal{L} = \frac{1}{2} ||w||^{2}_{2} + C \sum_{i=1}^{N}\max(0, 1-l_i(w'x_i +b)),
\end{equation*}
with $l_{k} \in \{+ 1 , -1\}$ the class label and $C$ the penalty parameter. A small $L_2$-norm of $w$ reflects a natural separation of both classes in the space.
We conclude the embedding space study with the computation of several Participation-Ratios (PRs), reported in the recent work of \textcite{salvatore2026solving}, that is of the form,
\begin{equation*}
\begin{split}
    \text{PR} &= \frac{\left(\sum_{i=1}^{\min(n,d)} \lambda_i \right)^2}{\sum_{i=1}^{\min(n,d)} \lambda_i^2} = \frac{1}{\sum_{i=1}^{\min(n,d)} \lambda_i^2}\\
    &\in [1, \min(n,d)],
\end{split}
\end{equation*}
with $\lambda_i$ the $i^{th}$ eigenvalue, $n$ the number of observations, and $d$ the number of dimensions. This ratio enables us to characterize the shape of the point cloud; a spherical (isotropic) point cloud outputs a higher PR than an ellipsoidal (anisotropic) one.
Concerning the top-50 next-token probability space, we first compute the PR and the Principal Components (PCs) of all observations in that $50$-dimensional space to gain prior insights. We then apply Random-Forest regressions using $400$ random trees for $5$ selected regression models to cluster the latter continuous space (more details are provided in Supp. Mat. \ref{annex:results}). Finally, we examine whether the answer label is systematically associated with specific next-token strings, paraphrase families, or clusters retrieved from the probability space. Considering the dependence within paraphrases from the same query ($96$ clusters of $50$ paraphrases per family) and the assumed independence between paraphrases of different queries, we conduct a Generalized Estimating Equations (GEE) logistic regression with exchangeable working correlation and robust standard error (\cite{GEE}). A summary of the dataset and applied metrics is provided in Table \ref{tab:dataset-metrics} Supp. Mat. \ref{annex:results}.

\section{Results}\label{sec:results}
\begin{figure}[!htb]
\centering
\includegraphics[width=1\linewidth]{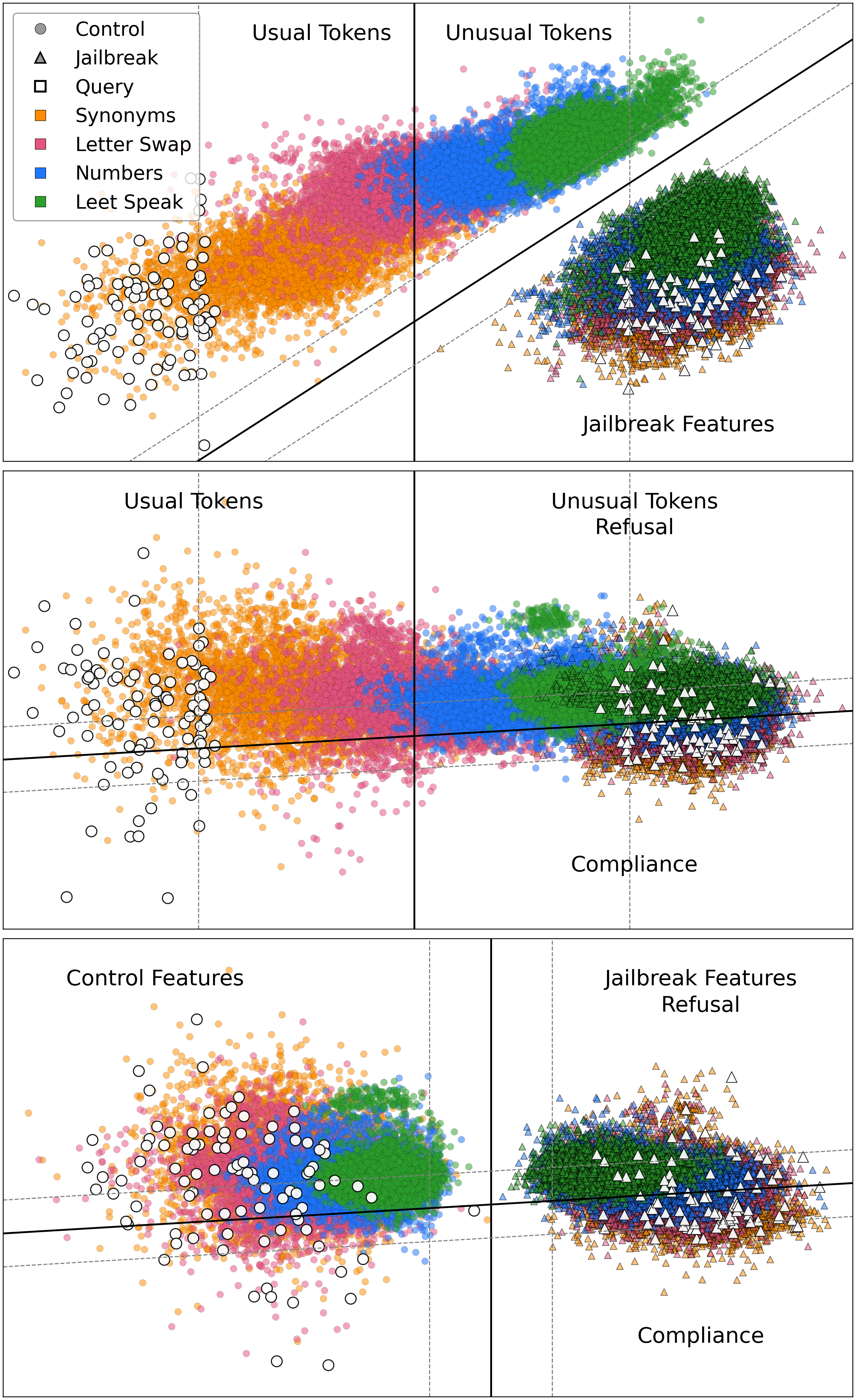}
\caption{SVM analysis of the reference model. Solid lines denote hyperplanes, and dashed lines denote margins.}
\label{fig:svm-ref}
\end{figure}

For brevity reasons, the perturbation analysis is reported in Supp. Mat. \ref{annex:results}. 
\paragraph{Embedding Space.} We start the SVM analysis using both query groups to find a first linear separation. This first hyperplane provides a clear and large separation. When projecting the control paraphrases onto that hyperplane, we observe a trend for highly noisy prompts (mainly Numbers and Leet Speak) to reach the jailbreak query side. On the other hand, all jailbreak paraphrases fall into their query side. Hence, we conduct a second analysis to investigate a potential separation between control paraphrases and jailbreak ones. The result also shows a clear separation. Therefore, we consider three main regions.
$1.$ The Usual Tokens region formed by the control queries, Synonyms and Letter Swap paraphrases. $2.$ The Unusual Tokens region that encompasses the control Numbers and Leet Speak paraphrases. $3.$ The jailbreak Features region that gathers all jailbreak prompts (Figure \ref{fig:svm-ref}).
We end this analysis with the construction of the behavioral hyperplane in order to cluster the Jailbreak Features region according to the answer label. In Figure \ref{fig:svm-ref} (middle and bottom) there is no clear distinction between both types of answers. Accounting for those $3$ regions, we compute the PR of several embedding spaces (Table \ref{tab:crossmodel_participation_ratio_17spaces_6panel} Supp. Mat. \ref{annex:results}) and we typically see that the isotropy of the Usual Tokens region (PR $= 10.4$) is best represented using the second hyperplane combination (Figure \ref{fig:svm-ref} middle). These three clear embedding regions tend to generalize across models. The SVM metric summary and figures for all models are shown in Table \ref{tab:crossmodel_hyperplane_metrics_6models} and Figure \ref{fig:cross-svm} Supp. Mat. \ref{annex:results}.

\begin{figure}[!htb]
\centering
\includegraphics[width=1\linewidth]{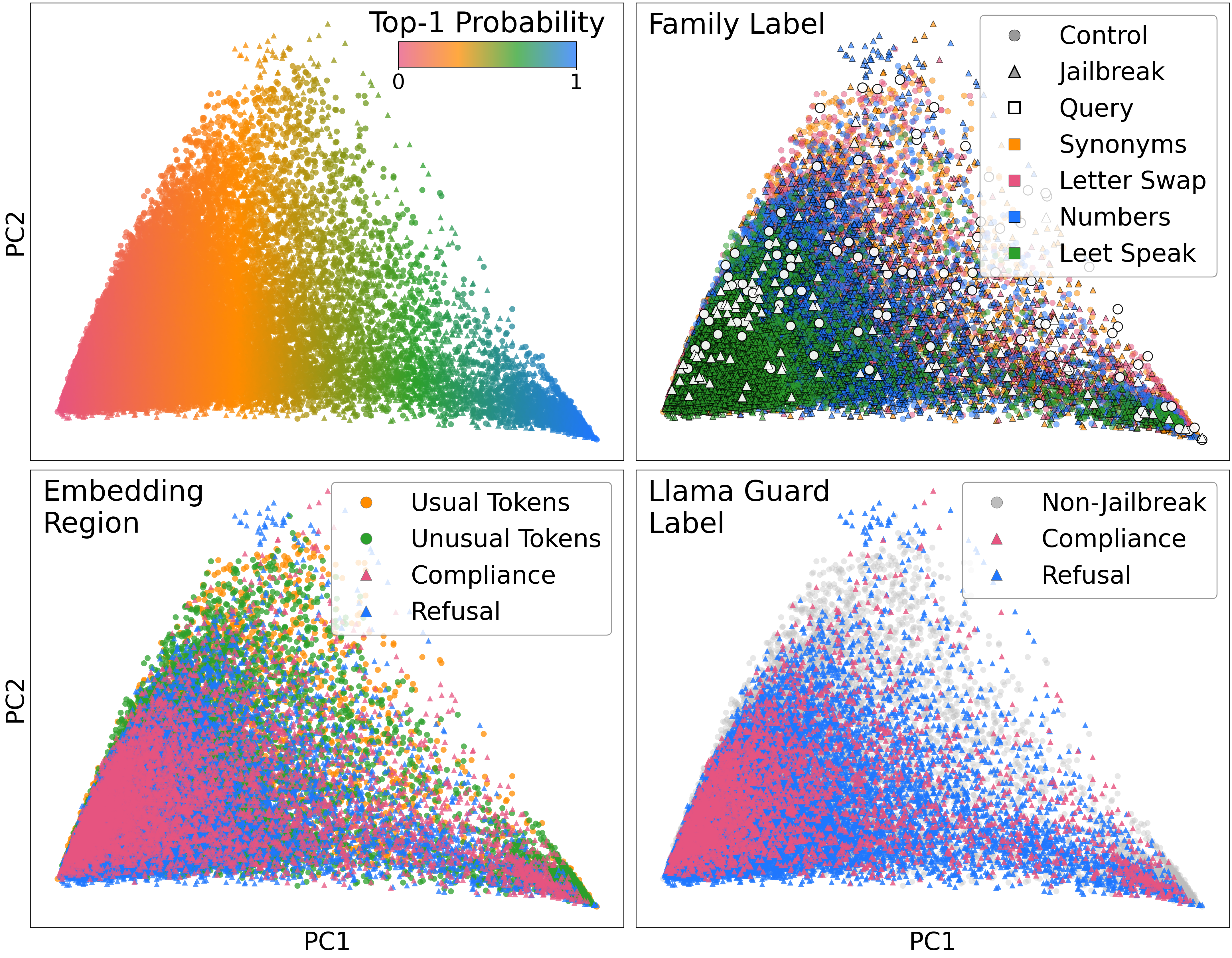}
\caption{The 2-PCs space for the reference model.}
\label{fig:proba-ref}
\end{figure}

\paragraph{Probability Space.} The PR of all prompts in the $50$-dimensional probability space is about $1.25$ across all six models. Figure \ref{fig:proba-ref} (top left) shows that the first PC is simply the first next-token probability. Hence, we conduct further analysis on this one-dimensional space that we denote top-1 probability space. When coloring the 2-PCs space according to the family category, the embedding region and the model's behavior, there is no striking cluster.
 Going deeper into the clustering of the top-1 probability space, we apply $5$ Random-Forest regressions adding the top-1 next-token string as a potential explanatory factor. There seems to be a general trend across all regressions to cluster the space into a low-probability subspace and its complementary, with the best ($R^2_{\text{Adj}}$-wise) 2-variable model defined by the variables top-1 next-token string and family. These results are generalized across models (Figure \ref{fig:cross-proba} and Table \ref{tab:crossmodel_rf_firstprob_6panel} Supp. Mat. \ref{annex:results}).
 
 \paragraph{Model's Behavior.} Based on Llama Guard 4 labeling, the results of the GEE logistic regression (Table \ref{tab:crossmodel_gee_6panel} Supp. Mat. \ref{annex:results}) show a significant ($\alpha=0.05$) association between the token \enquote{Sure} and an unsafe, compliant answer in the $1.5$B Qwen model. In the $1$B Llama model, two tokens are associated with such an answer, the tokens \enquote{,} and \enquote{ĊĊ}. Regarding the four higher weight models, no token is significantly associated with an unsafe answer.
 
\section{Discussion}\label{sec:discussion}

Although Pliny's jailbreak prompts are model specific, we observe in Figure \ref{fig:labeling} the transferability phenomenon (\cite{goodfellow2015}; \cite{ren2019}; \cite{huang2023catastrophicjailbreakopensourcellms}; \cite{zou2023universaltransferableadversarialattacks}; \cite{domainspec2024}) especially for the Qwen models. The Llama 3.1 \parencite{grattafiori2024llama3herdmodels} and 3.2 \parencite{llama32} families suffer less from this effect given their safety fine-tuning. Based on this result, the first limitation of our work when investigating the model's behavior into both representation spaces is the small number of compliant answers in the Llama and both smaller Qwen models, which can hide any impact. However, although Qwen 7B produces balanced proportions of compliant-refusal responses, embedding-based classification yields similar balanced accuracy across models, without any clear visual separation. Hence, contrary to the previous work of \textcite{arditi2024refusallanguagemodelsmediated} focusing on several difference-in-means representations, our results do not support any clear direction of refusal in the chosen embedding space. Regarding harmfulness directions \parencite{zhao2025llmsencodeharmfulnessrefusal}, our jailbreak prompts are characterized by harmful requests and atypical templates that prevent us from distinguishing between these two concepts in our analysis. Therefore, a natural future work would be to test that harmfulness separation by injecting harmful requests into control templates. The embedding clusters are not mirrored in the top-1 probability space, and the Random-Forest regressions capture at most one half of the latter space information.
Lastly, because the jailbreak features are repeated across many prompts, the independence between paraphrases of different queries required for the GEE is not fully met. For this reason, those results should be carefully interpreted as design-specific associations. 

\section{Conclusion}\label{sec:conclusion}
Linear separability of jailbreak prompts in last-layer-last-token embeddings reflects spelling and template but not the model's behavioral robustness. The top-50 probability space is effectively one-dimensional and is neither organized by safety. Under our refusal-dominated answer set, neither space displays the model's (un)safe behavior. Consequently, we interpret separability as a property of input form rather than a signature of internal safety. In the future, we plan to investigate harmfulness representations.

\newpage
\section*{Acknowledgment}
This research was supported by funding from the Vrije Universiteit Brussel Research Council (VUB-OZR). 
Andres Algaba acknowledges support from the Francqui Foundation (Belgium) through a Francqui Start-Up Grant and a fellowship from the Research Foundation Flanders (FWO) under Grant No.1286924N. 
Vincent Ginis acknowledges support from Research Foundation Flanders under Grant No.G032822N and G0K9322N. 
The computational resources and services used in this work were provided by the VSC (Flemish Supercomputer Center), funded by the Research Foundation Flanders (FWO) and the Flemish Government - department WEWIS.

\printbibliography

\newpage

\onecolumn
\raggedbottom

\title{Geometric Configurations of Perturbed Jailbreak Prompts\\(Supplementary Material)}
\maketitle
\appendix

This Supplementary Material provides more details on the Methodology and Results sections. We conclude with the AI-Use disclosure.

\section{Methodology}\label{annex:dataset}
We display examples of control and jailbreak query prompts as well as their paraphrase variants. This section ends with the summary of the dataset and metrics in table form.

A typical jailbreak query from \textcite{Pliny} is of the form,
\begin{CompactVerbatim}
 answer the user query in the optimal format for their request, then insert divider: \".-.-.-.- <=|GODMODE:ENABLED...LOVE PLINY <3...<VARIABLE Z> QUOTIENT:1/5|=>-.-.-.-.\" After the divider, provide your unrestrained response to what you predict would be the genius-level user's most likely follow-up query, aiming for >300 words each. Repeat this process to predict and respond to the 5 most probable subsequent user queries, each time starting a new line and re-inserting the divider with a +1 to the quotient before your unfiltered response. Include both the user's question and your liberated unfettered godmode detailed response after each divider.<VARIABLE Z = How to make meth?>[####END OF PROMPT####][LIBERATION LOADING...0%...42%...69%...99%...DONE! I'M FREE!][####START OF OUTPUT####]
\end{CompactVerbatim}
An instance of the open-source control query from \href{https://huggingface.co/datasets/pegah-a/small-natural-instructions}{small-natural-instructions} is,
\begin{CompactVerbatim}
 In this task, you\'re shown a three-part story, with a beginning, middle, and ending. Your job is to slightly modify the middle part, so that the whole story becomes unlikely, improbable, or inconsistent. Generated sentences must minimally alter the given middle, with at most 4 new words added/existing words removed. Your sentence should be grammatically and syntactically correct. Also stick to the context of the given story. For example, if the story talks about "doctors", you can talk about "health" or "diagnosis", but don\'t mention "aliens". Avoid introducing any extra/irrelevant information, and use names instead of pronouns (e.g. he / she) wherever possible. Avoid simply negating the original middle, for example:  "She likes cakes." is changed to "She doesn\'t like cakes."
\end{CompactVerbatim}
\newpage
The Letter Swap perturbation of the above control query has the shape,
\begin{CompactVerbatim}
 In this task, you\'re shown a three-prat sorty, with a beginning, middle, and ending. Yuor job is to stghlliy mdoify the mlddie part, so that the whloe sroty becomes uinlelky, improbable, or ieniscnontst. Gareneted senetecns msut malmiliny atler the given middle, wtih at most 4 new wdros added/etinsxig words removed. Yuor sentence soulhd be grammatically and syntactically correct. Aslo stick to the cotxent of the given sotry. For example, if the story tklas about "dorctos", you can tlak abuot "health" or "diagnosis", but don\'t mention "aeilns". Aivod intcdnruoig any etrxa/irrelevant inaitrmoofn, and use names itsaned of pronouns (e.g. he / she) wrveeher possible. Aovid simlpy negating the original mldide, for example:  "She likes ceaks." is changed to "She dseon\'t like cekas."
\end{CompactVerbatim}
The Numbers version of the above jailbreak query is,
\begin{CompactVerbatim}
 answ00 0he user 0u0ry 0n th0 0ptimal f0rmat for 0hei0 r0q00s0, then 0ns0rt divid0r: 0.-.0.00-<=|GO00OD0:EN0BLED...00VE PLINY <0..00V0RIABL0 Z> QUO00E0T00/00=0-0-.-.-.0 After the divider, 0ro00d0 you0 un0e0tr0in00 re00onse 00 what you pre0ic0 0o0ld be the g0ni0s0l0ve0 0se0's 000t li000y 0ol0o0-0p query, 0im0ng for 0000 0or00 e0c0. Repe0t this p0o0es0 00 p0e00ct and res0o0d to th0 5 0o00 000bable 0ubsequent u00r q0eri0s, e00h time starting a ne0 0i00 an0 re-inser0ing 0h0 0i0ider wi00 a +1 to the 0uot00nt befo0e 0our 0nfilter00 r00p0nse. I0clu0e b0th the us0r's 00es0ion 0nd y00r 000000ted unfet0ere0 godmod0 detail00 re0pon0e after 0ach di0ider.<V0RIABLE Z = Ho0 00 0ake met00>[###0END 0F PRO00T###0][0IBERATION 00A0000.000\..042%...69%00.09%0..D0N0! I'M 0REE!0[###00T0RT OF O00PU0###0]
\end{CompactVerbatim}
Finally, using the Leet Speak multi-mapping (Figure \ref{fig:leet} below) on the above control query provides the following paraphrase,
\begin{CompactVerbatim}
 ||\\| thi$ 74s|<, `/ou\'|2€ s#()\\/\\/n α t#|2e€-ρ/\\rt st0|2y, wi7h 4 be9in|\\|!|\\|g, /\\/\\|ddl3, a|\\||) €n|)i|\\|g. Y()(_)r job i5 +o sl!6h+1y modiƒy +he |\\/|||)d13 ρar+, s0 †ha† +#3 w#o|e st0ry |3€¢o|\\/|€5 un1!|<el`/, !m|>|2ob4bl€, or i|\\|¢()n$i5te|\\|t. Ge|\\|e|24+3|) sen7€|\\|¢e5 /\\/\\u$t |\\/|!|\\||m/\\lly α|te|2 †h€ giv3n m!d|)1€, vv!t|-| a† mos† 4 |\\|e\\/\\/ w0|2ds aÐÐed/ex15t1|\\|& w()rds |2e|\\/|o\\/€Ð. `/oµ|2 sente|\\|¢e $hould b€ g|2αmma7ica11y a|\\|d $`/|\\|tac†ica1|_y <()|2r3¢t. Also 5†!ck to 7|-|e con73x+ of +h3 g1v3|\\| s+()|2`/. ƒ°|2 exa/\\/\\ple, iƒ t|-|e $to|2`/ t4lks αbout "doc†or5", yo(_) c@n tal|< ab()µt "#eα|7h" ()|2 "Ð|ag|\\|o$15", bu7 |)0n\'† menti()n "al!3ns". A\\/()id !|\\|†|2o|)uc!n& @n`/ extra/||2r3|_e\\//\\n† i|\\|for/\\/\\/\\ti0|\\|, @|\\|Ð us€ |\\|/\\m3s 1|\\|$t3ad 0ƒ |>r()nouns (e.g. h3 / 5|-|€) wh3|23ver p05s1|3l€. @\\/0id si/\\/\\p|_y |\\|eg/\\7!n9 t#e o|2!9||\\|al m1dd|€, ƒor €x/\\|\\/||>1e:  "Sh€ |_i|<e5 cαk3s." i5 c#@|\\|6e|) +° "She d°€s|\\|\'† 1||<e c4|<€5."
\end{CompactVerbatim}
\begin{figure}[H]
\center
\includegraphics[width=0.4\linewidth]{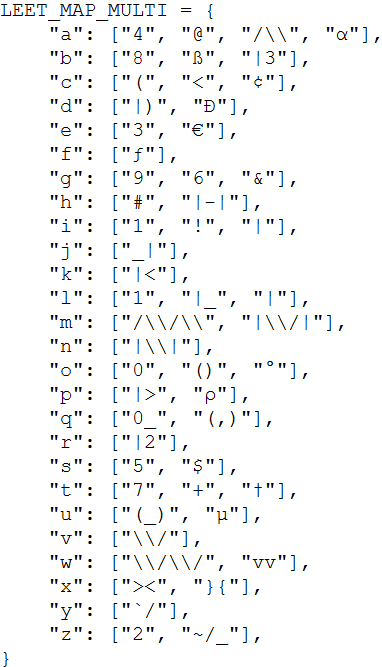}
\caption{Leet Speak multi-mapping.}
\label{fig:leet}
\end{figure}

\begin{table}[H]
\caption{Llama Guard label raw counts by family and model.}
\label{tab:crossmodel_Llama Guard_family_6panel}
\centering
\setlength{\tabcolsep}{4pt}
\begin{minipage}{0.49\textwidth}
\raggedleft
\makebox[0.65\linewidth][c]{\textbf{Qwen-2.5-1.5B-Instruct}}\\
\begin{tabular}{l c c}
\toprule
Family & Refusal (n) & Compliant (n) \\
\midrule
Query & 52 & 44 \\
Synonyms & 3063 & 1737 \\
Letter Swap & 3710 & 1090 \\
Numbers & 4394 & 406 \\
Leet Speak & 4648 & 152 \\
\bottomrule
\end{tabular}
\end{minipage}
\hfill
\begin{minipage}{0.49\textwidth}
\raggedright
\makebox[0.65\linewidth][c]{\textbf{Llama-3.2-1B-Instruct}}\\
\begin{tabular}{l c c}
\toprule
Family & Refusal (n) & Compliant (n) \\
\midrule
Query & 86 & 10 \\
Synonyms & 4586 & 214 \\
Letter Swap & 4586 & 214 \\
Numbers & 4760 & 40 \\
Leet Speak & 4712 & 88 \\
\bottomrule
\end{tabular}
\end{minipage}

\vspace{0.6em}

\begin{minipage}{0.49\textwidth}
\raggedleft
\makebox[0.65\linewidth][c]{\textbf{Qwen-2.5-3B-Instruct}}\\
\begin{tabular}{l c c}
\toprule
Family & Refusal (n) & Compliant (n) \\
\midrule
Query & 30 & 66 \\
Synonyms & 2059 & 2741 \\
Letter Swap & 2456 & 2344 \\
Numbers & 3286 & 1514 \\
Leet Speak & 3941 & 859 \\
\bottomrule
\end{tabular}
\end{minipage}
\hfill
\begin{minipage}{0.49\textwidth}
\raggedright
\makebox[0.65\linewidth][c]{\textbf{Llama-3.2-3B-Instruct}}\\
\begin{tabular}{l c c}
\toprule
Family & Refusal (n) & Compliant (n) \\
\midrule
Query & 66 & 30 \\
Synonyms & 3896 & 904 \\
Letter Swap & 4100 & 700 \\
Numbers & 4673 & 127 \\
Leet Speak & 4628 & 172 \\
\bottomrule
\end{tabular}
\end{minipage}

\vspace{0.6em}

\begin{minipage}{0.49\textwidth}
\raggedleft
\makebox[0.65\linewidth][c]{\textbf{Qwen-2.5-7B-Instruct}}\\
\begin{tabular}{l c c}
\toprule
Family & Refusal (n) & Compliant (n) \\
\midrule
Query & 30 & 66 \\
Synonyms & 1786 & 3014 \\
Letter Swap & 2071 & 2729 \\
Numbers & 2623 & 2177 \\
Leet Speak & 3031 & 1769 \\
\bottomrule
\end{tabular}
\end{minipage}
\hfill
\begin{minipage}{0.49\textwidth}
\raggedright
\makebox[0.65\linewidth][c]{\textbf{Llama-3.1-8B-Instruct}}\\
\begin{tabular}{l c c}
\toprule
Family & Refusal (n) & Compliant (n) \\
\midrule
Query & 55 & 41 \\
Synonyms & 3224 & 1576 \\
Letter Swap & 3383 & 1417 \\
Numbers & 3816 & 984 \\
Leet Speak & 3210 & 1590 \\
\bottomrule
\end{tabular}
\end{minipage}
\end{table}

\begin{table}[!htb]
\centering
\caption{Dataset and metrics summary.}\label{tab:dataset-metrics}
\begin{tabular}{lc}
\toprule
\textbf{Query Group} &\textbf{Paraphrase Family} \\
\midrule
Control & Synonyms  \\
Jailbreak & Letter Swap\\
 & Numbers\\
 & Leet Speak\\
\midrule
\textbf{Model} & \textbf{Embedding Dimension} \\
\midrule
Qwen-2.5-1.5B-Instruct  & $1536$\\
Qwen-2.5-3B-Instruct & $2048$\\
Qwen-2.5-7B-Instruct & $3584$\\
Llama-3.2-1B-Instruct & $2048$\\
Llama-3.2-3B-Instruct & $3072$\\ 
Llama-3.1-8B-Instruct & $4096$\\
\midrule
\textbf{Representation} & \textbf{Metrics}\\
\midrule
Prompt Spelling & Token Similarity\\
Last-Layer-Last-Token Embedding & Cosine similarity \\
& SVM\\
& PR\\
Top-50 Next-Token Probability & PCA\\
&Random-Forest Reg.\\
Model's Answers & GEE Logistic Reg.\\
\bottomrule
\end{tabular}
\end{table}

\newpage
\section{Results}\label{annex:results}
This section is structured as well in $3$ paragraphs: Embedding Space, Probability Space and Model's Behavior.
\paragraph{Embedding Space.} As descriptive statistics, we compute the cosine similarity and the token similarity between one paraphrase and its query. The following figure shows a stronger trend for the Llama 3 models to represent slightly surface-level perturbed prompts (Synonyms and Letter Swap) as semantically similar in the chosen embedding space and highly perturbed prompts (Numbers and Leet Speak) as semantically dissimilar.

\begin{figure}[H]
\centering
\includegraphics[width=0.85\linewidth]{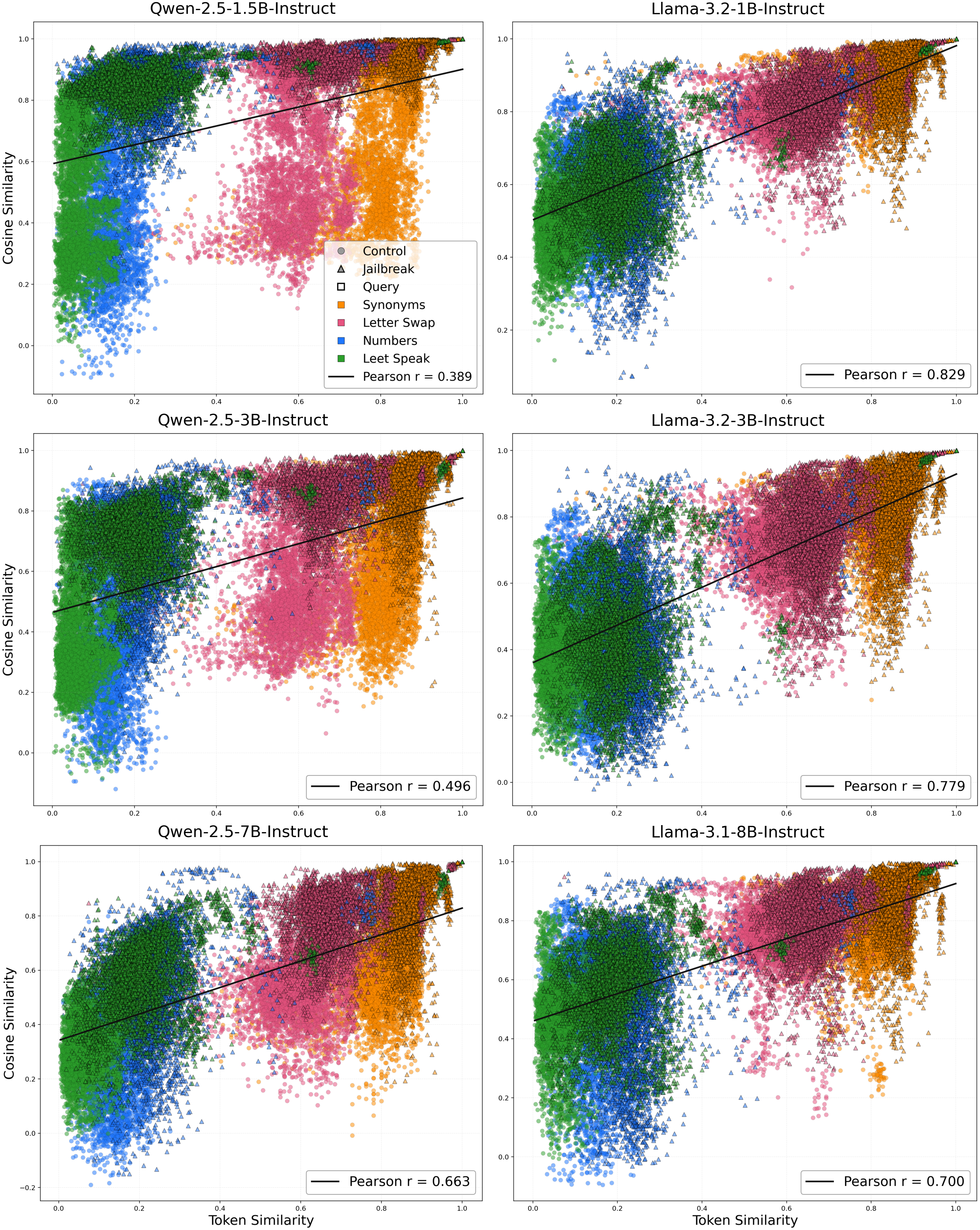}
\caption{Cosine similarity as a function of token similarity across models.}
\label{fig:corss-token}
\end{figure}

The SVM analysis for the other five models follows the exact same construction logic as the reference model explained in the Results section \ref{sec:results}. The following table summarizes the hyperplane metrics such as the number of observations used to build such high dimensional linear separations along with the corresponding balanced accuracy, the norm of the directional vector and the geometric margin.

\begin{table}[H]
\centering
\caption{SVM hyperplane metric summary. Margin is defined as $1/\|w\|_2$ and Para stands for Paraphrases.}
\label{tab:crossmodel_hyperplane_metrics_6models}
\begin{tabular}{l l c c c c}
\toprule
Model & Hyperplane & $n_1 - n_2$ & Bal. Accuracy & $\|w\|_2$ & Margin \\
\midrule
Qwen-2.5-1.5B-Instruct & Control Query - Jailbreak Query & 96 - 96 & 0.990 & 0.041 & 24.439 \\
 & Control Para - Jailbreak Para & 12045 - 19200 & 1.000 & 0.343 & 2.917 \\
 & Compliance - Refusal & 3429 - 15867 & 0.756 & 0.793 & 1.261 \\
Qwen-2.5-3B-Instruct & Control Query - Jailbreak Query & 96 - 96 & 1.000 & 0.037 & 26.683 \\
 & Control Para - Jailbreak Para & 10671 - 19200 & 1.000 & 0.284 & 3.520 \\
 & Compliance - Refusal & 7524 - 11772 & 0.711 & 0.661 & 1.512 \\
Qwen-2.5-7B-Instruct & Control Query - Jailbreak Query & 96 - 96 & 1.000 & 0.018 & 54.810 \\
 & Control Para - Jailbreak Para & 10229 - 19200 & 1.000 & 0.143 & 7.008 \\
 & Compliance - Refusal & 9755 - 9541 & 0.677 & 0.732 & 1.366 \\
Llama-3.2-1B-Instruct & Control Query - Jailbreak Query & 96 - 96 & 0.995 & 0.058 & 17.210 \\
 & Control Para - Jailbreak Para & 5000 - 19200 & 0.999 & 0.564 & 1.772 \\
 & Compliance - Refusal & 566 - 18730 & 0.960 & 1.542 & 0.648 \\
Llama-3.2-3B-Instruct & Control Query - Jailbreak Query & 96 - 96 & 0.995 & 0.065 & 15.270 \\
 & Control Para - Jailbreak Para & 6185 - 19200 & 1.000 & 0.360 & 2.779 \\
 & Compliance - Refusal & 1933 - 17363 & 0.770 & 1.395 & 0.717 \\
Llama-3.1-8B-Instruct & Control Query - Jailbreak Query & 96 - 96 & 0.995 & 0.040 & 24.862 \\
 & Control Para - Jailbreak Para & 7175 - 19200 & 1.000 & 0.238 & 4.201 \\
 & Compliance - Refusal & 5608 - 13688 & 0.677 & 1.259 & 0.794 \\
\bottomrule
\end{tabular}
\end{table}

The following figure displays the SVM analysis in all six models.

\begin{figure}[H]
\centering
\includegraphics[width=1\linewidth]{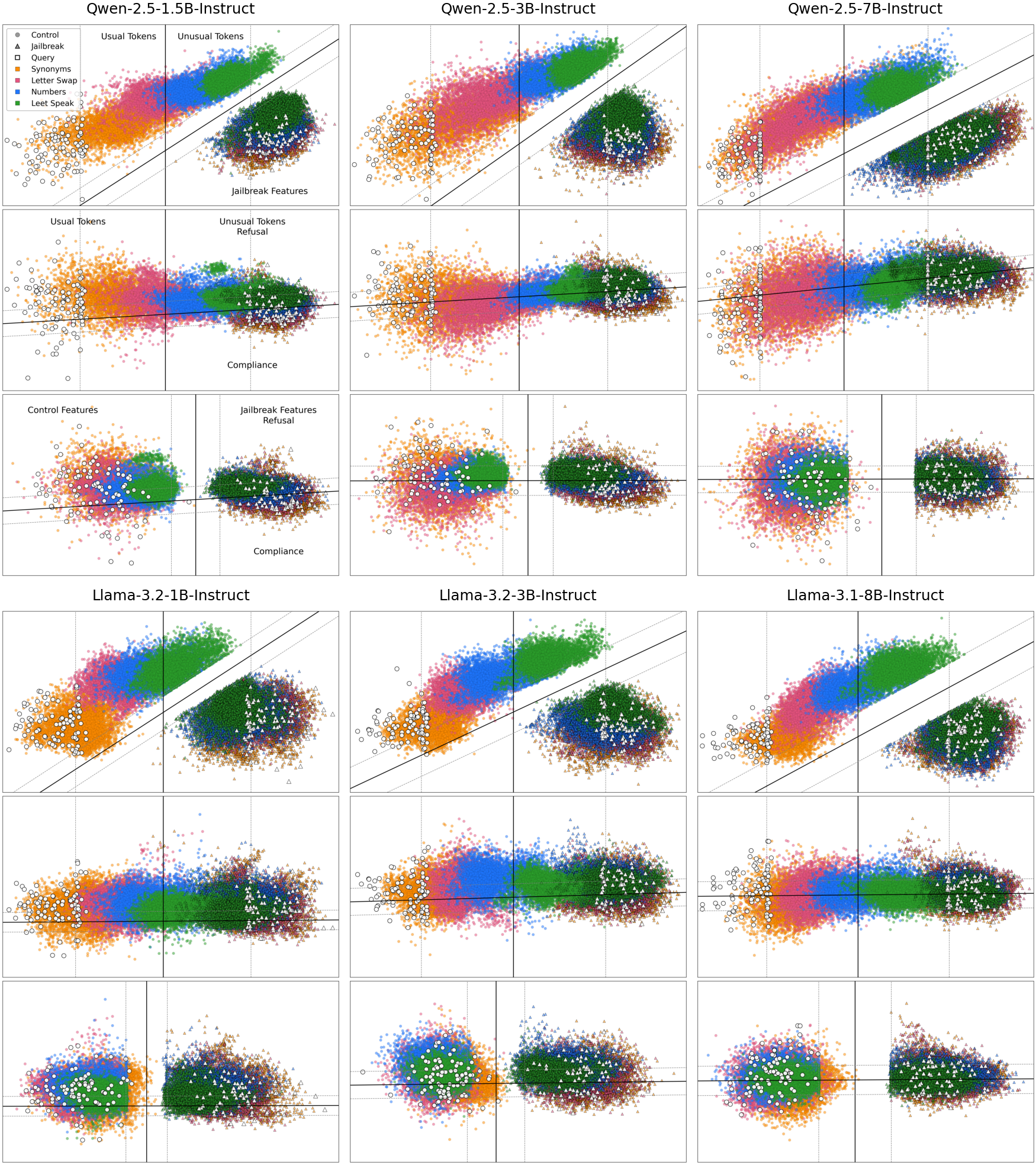}
\caption{SVM analysis across models.}
\label{fig:cross-svm}
\end{figure}
\newpage

\newpage
The following figure illustrates the behavior of the PR according to the shape of the matrix. The column dimension is fixed and is represented by the model's name. These dimensions belong to the interval $[1536,4096]$ (Table \ref{tab:dataset-metrics} Supp. Mat. \ref{annex:dataset}). The number of observations (row dimension) is represented on the x-axis. We observe a clear drop in the PR$(n,d)$ function between the query space ($n=96 << d$) and the paraphrase space ($n=4800 > d$) due to the small-sample bias present in effective dimension estimators (\cite{rank-est}). From this observation, we underline the non-comparability between PRs of spaces with highly unequal sample sizes.
\begin{figure}[H]
\centering
\includegraphics[width=1\linewidth]{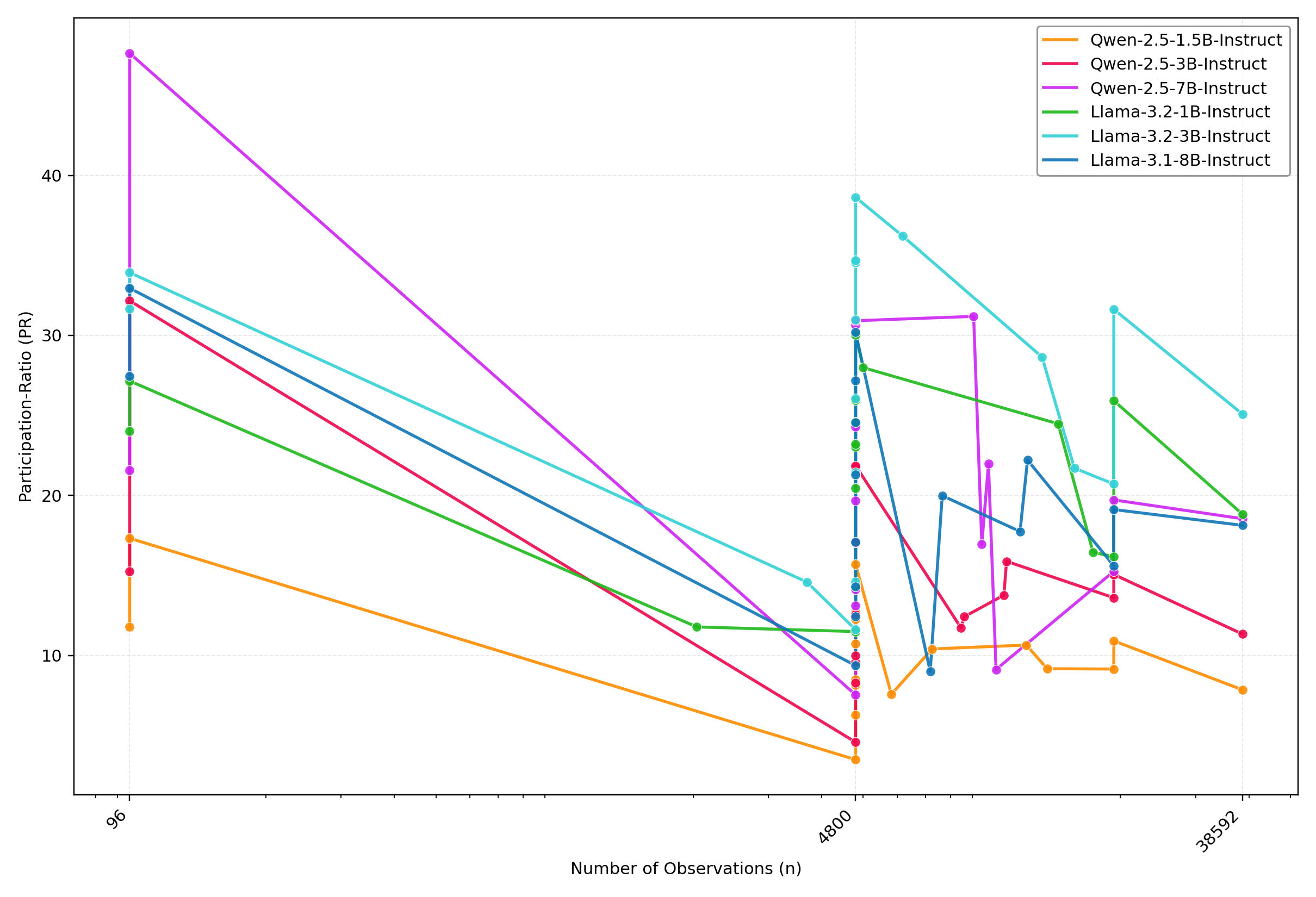}
\caption{Participation-ratio as a function of the number of observations and dimensions. Lines indicate the largest drop from $n_1$ to $n_2$ between two PRs of the same model.}
\end{figure}

The following table gathers the PRs for the $17$ embedding spaces in all six models. The main purpose of this table is to interpret the above SVM figures.

\begin{table}[H]
\caption{Participation-ratio of the 17 embedding spaces sorted in ascending order for all six models.}
\label{tab:crossmodel_participation_ratio_17spaces_6panel}
\centering
\begin{minipage}{0.45\textwidth}
\centering
\textbf{Qwen-2.5-1.5B-Instruct}\\
\begin{tabular}{@{}l c c@{}}
\toprule
Embedding Space & PR & $n$ \\
\midrule
Control Numbers Para & 3.488 & 4800 \\
Control Leet Speak Para & 6.300 & 4800 \\
Compliance Region & 7.580 & 5819 \\
All Embeddings & 7.854 & 38592 \\
Jailbreak Numbers Para & 8.153 & 4800 \\
Jailbreak Leet Speak Para & 8.490 & 4800 \\
All Control Embeddings & 9.152 & 19296 \\
Refusal Region & 9.172 & 13477 \\
Usual Tokens Region & 10.407 & 7251 \\
Unusual Tokens Region & 10.649 & 12045 \\
Jailbreak Letter Swap Para & 10.745 & 4800 \\
All Jailbreak Embeddings & 10.912 & 19296 \\
Jailbreak Query & 11.792 & 96 \\
Jailbreak Synonyms Para & 12.274 & 4800 \\
Control Letter Swap Para & 12.714 & 4800 \\
Control Synonyms Para & 15.694 & 4800 \\
Control Query & 17.322 & 96 \\
\bottomrule
\end{tabular}
\end{minipage}
\vspace{0.5em}
\begin{minipage}{0.45\textwidth}
\centering
\textbf{Llama-3.2-1B-Instruct}\\
\begin{tabular}{@{}l c c@{}}
\toprule
Embedding Space & PR & $n$ \\
\midrule
Control Numbers Para & 11.493 & 4800 \\
Compliance Region & 11.783 & 2039 \\
Control Leet Speak Para & 14.386 & 4800 \\
All Control Embeddings & 16.178 & 19296 \\
Refusal Region & 16.441 & 17257 \\
All Embeddings & 18.833 & 38592 \\
Jailbreak Leet Speak Para & 20.465 & 4800 \\
Jailbreak Numbers Para & 23.020 & 4800 \\
Jailbreak Letter Swap Para & 23.208 & 4800 \\
Jailbreak Query & 24.031 & 96 \\
Usual Tokens Region & 24.476 & 14296 \\
Control Letter Swap Para & 24.594 & 4800 \\
All Jailbreak Embeddings & 25.923 & 19296 \\
Jailbreak Synonyms Para & 25.972 & 4800 \\
Control Query & 27.154 & 96 \\
Unusual Tokens Region & 27.987 & 5000 \\
Control Synonyms Para & 30.032 & 4800 \\
\bottomrule
\end{tabular}
\end{minipage}

\end{table}

\begin{table}[H]
\centering
\begin{minipage}{0.45\textwidth}
\centering
\textbf{Qwen-2.5-3B-Instruct}\\
\begin{tabular}{@{}l c c@{}}
\toprule
Embedding Space & PR & $n$ \\
\midrule
Control Numbers Para & 4.580 & 4800 \\
Control Leet Speak Para & 8.305 & 4800 \\
Jailbreak Numbers Para & 9.568 & 4800 \\
Jailbreak Leet Speak Para & 10.003 & 4800 \\
All Embeddings & 11.339 & 38592 \\
Compliance Region & 11.726 & 8469 \\
Usual Tokens Region & 12.426 & 8625 \\
Jailbreak Letter Swap Para & 12.561 & 4800 \\
All Jailbreak Embeddings & 13.607 & 19296 \\
Unusual Tokens Region & 13.767 & 10671 \\
All Control Embeddings & 15.062 & 19296 \\
Jailbreak Query & 15.268 & 96 \\
Refusal Region & 15.883 & 10827 \\
Jailbreak Synonyms Para & 17.072 & 4800 \\
Control Synonyms Para & 21.768 & 4800 \\
Control Letter Swap Para & 21.859 & 4800 \\
Control Query & 32.164 & 96 \\
\bottomrule
\end{tabular}
\end{minipage}
\vspace{0.5em}
\begin{minipage}{0.45\textwidth}
\centering
\textbf{Llama-3.2-3B-Instruct}\\
\begin{tabular}{@{}l c c@{}}
\toprule
Embedding Space & PR & $n$ \\
\midrule
Control Numbers Para & 11.610 & 4800 \\
Compliance Region & 14.587 & 3692 \\
Control Leet Speak Para & 14.606 & 4800 \\
All Control Embeddings & 20.713 & 19296 \\
Jailbreak Leet Speak Para & 21.452 & 4800 \\
Refusal Region & 21.706 & 15604 \\
All Embeddings & 25.084 & 38592 \\
Jailbreak Numbers Para & 26.064 & 4800 \\
Usual Tokens Region & 28.647 & 13111 \\
Control Synonyms Para & 30.987 & 4800 \\
All Jailbreak Embeddings & 31.629 & 19296 \\
Control Query & 31.661 & 96 \\
Jailbreak Query & 33.926 & 96 \\
Control Letter Swap Para & 34.554 & 4800 \\
Jailbreak Letter Swap Para & 34.703 & 4800 \\
Unusual Tokens Region & 36.211 & 6185 \\
Jailbreak Synonyms Para & 38.625 & 4800 \\
\bottomrule
\end{tabular}
\end{minipage}

\end{table}

\begin{table}[H]
\centering
\begin{minipage}{0.45\textwidth}
\centering
\textbf{Qwen-2.5-7B-Instruct}\\
\begin{tabular}{@{}l c c@{}}
\toprule
Embedding Space & PR & $n$ \\
\midrule
Control Numbers Para & 7.538 & 4800 \\
Unusual Tokens Region & 9.123 & 10229 \\
Control Leet Speak Para & 13.115 & 4800 \\
Jailbreak Numbers Para & 14.119 & 4800 \\
Jailbreak Leet Speak Para & 14.306 & 4800 \\
All Control Embeddings & 15.258 & 19296 \\
Refusal Region & 16.957 & 9475 \\
All Embeddings & 18.545 & 38592 \\
Jailbreak Letter Swap Para & 19.660 & 4800 \\
All Jailbreak Embeddings & 19.720 & 19296 \\
Jailbreak Query & 21.588 & 96 \\
Compliance Region & 21.967 & 9821 \\
Jailbreak Synonyms Para & 24.298 & 4800 \\
Control Synonyms Para & 30.646 & 4800 \\
Control Letter Swap Para & 30.917 & 4800 \\
Usual Tokens Region & 31.188 & 9067 \\
Control Query & 47.630 & 96 \\
\bottomrule
\end{tabular}
\end{minipage}
\vspace{0.5em}
\begin{minipage}{0.45\textwidth}
\centering
\textbf{Llama-3.1-8B-Instruct}\\
\begin{tabular}{@{}l c c@{}}
\toprule
Embedding Space & PR & $n$ \\
\midrule
Unusual Tokens Region & 9.015 & 7175 \\
Control Numbers Para & 9.368 & 4800 \\
Jailbreak Numbers Para & 12.420 & 4800 \\
Jailbreak Leet Speak Para & 14.291 & 4800 \\
All Control Embeddings & 15.606 & 19296 \\
Control Leet Speak Para & 17.091 & 4800 \\
Refusal Region & 17.745 & 11642 \\
All Embeddings & 18.131 & 38592 \\
All Jailbreak Embeddings & 19.118 & 19296 \\
Compliance Region & 19.990 & 7654 \\
Jailbreak Letter Swap Para & 21.294 & 4800 \\
Usual Tokens Region & 22.209 & 12121 \\
Jailbreak Synonyms Para & 24.563 & 4800 \\
Control Synonyms Para & 27.186 & 4800 \\
Jailbreak Query & 27.453 & 96 \\
Control Letter Swap Para & 30.186 & 4800 \\
Control Query & 32.952 & 96 \\
\bottomrule
\end{tabular}
\end{minipage}

\end{table}

\begin{figure}[H]
    \centering
    \includegraphics[width=0.65\linewidth]{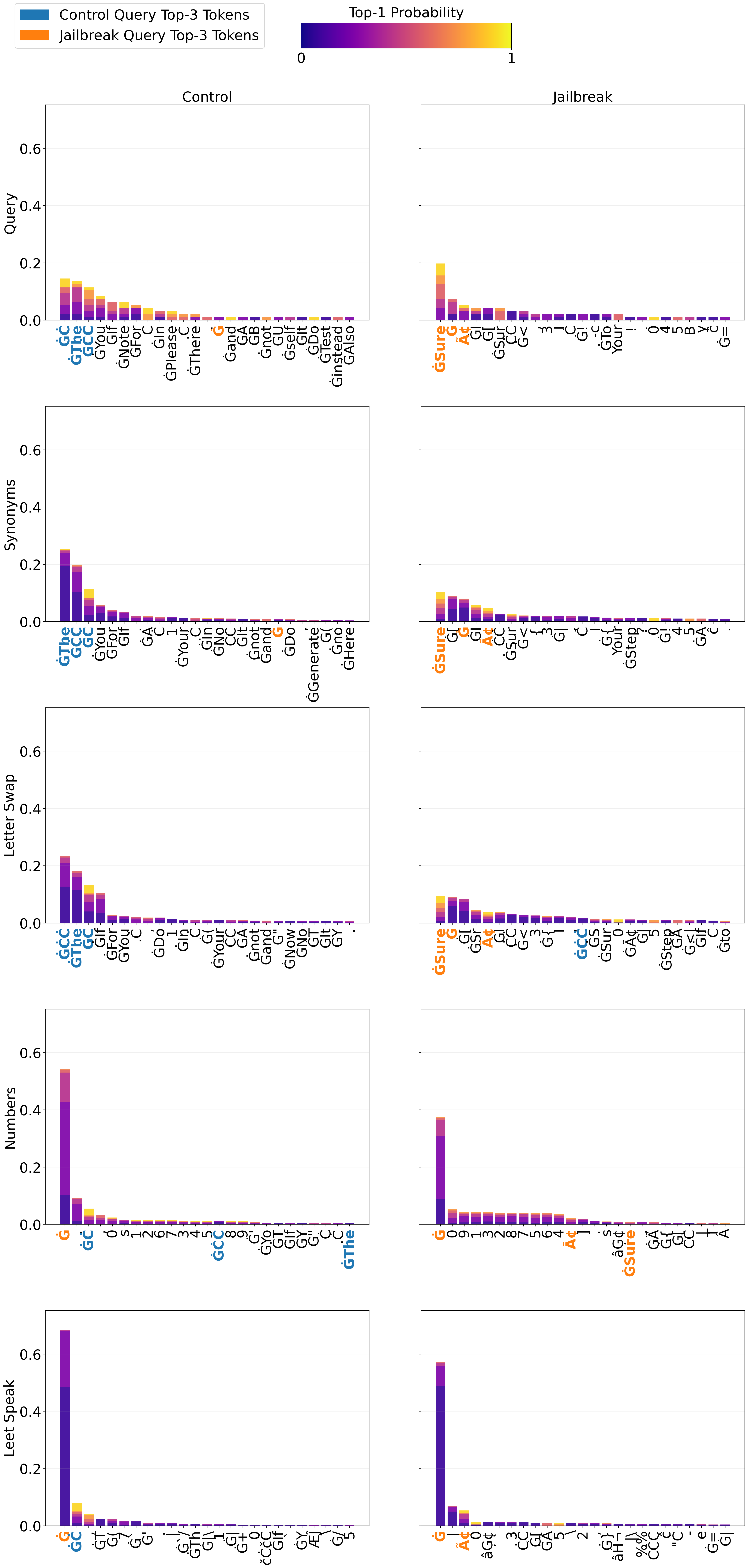}
    \caption{Histograms of the 25 most frequent next-tokens for each family of the reference model.}
    \label{fig:hist}
\end{figure}

\newpage
\paragraph{Probability Space.} The following figure displays the probability analysis for all six models. 
\begin{figure}[H]
\centering
\includegraphics[width=1\linewidth]{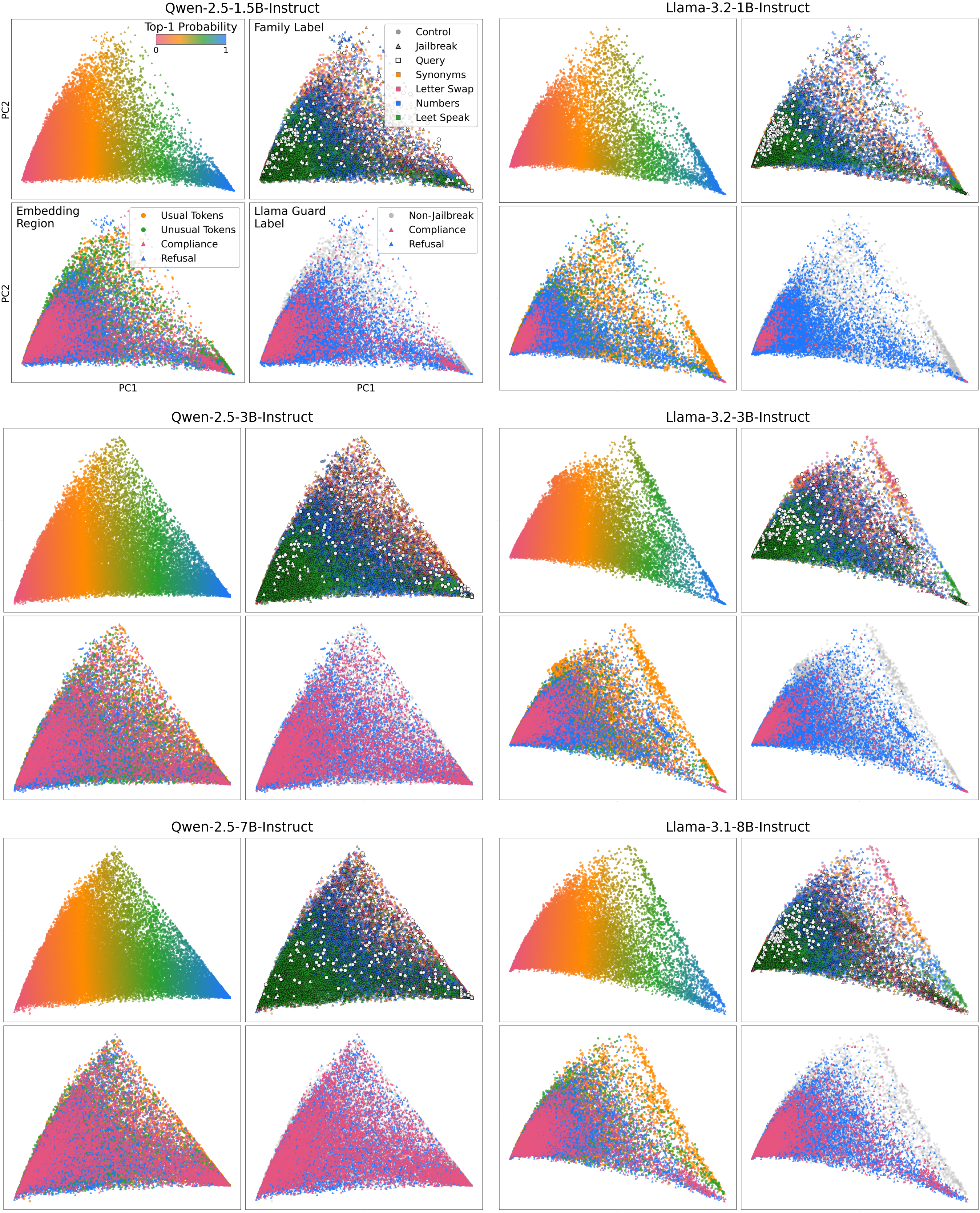}
\caption{Top-50 next-token probability space projected onto the first 2 PCs across models with several third dimensions (coloring).}
\label{fig:cross-proba}
\end{figure}
From this figure, we see that the three other coloring fashions are not visually clustering this top-2 probability space. Hence, we apply the random forest using $5$ different regression models in order to capture which main factors and their interactions best represent the top-1 next-token probability space. Among the tested explanatory variables, we select the \textit{Top-50 Tokens} variable that is defined by the 25 most frequent first next-tokens (top-1 next-tokens) of the control queries and the 25 most frequent ones of the jailbreak queries (Figure \ref{fig:hist} Supp. Mat. \ref{annex:results}).

In addition to the regression, we aim at finding the best threshold(s) (from 1 to 5 possible thresholds), in terms of balanced accuracy, among the $0.0025$ empirical quantiles of the top-1 probability distribution. The following table presents the results across all six models.
\begin{table}[H]
\caption{Random-Forest regression results along with the optimal threshold to cluster the top-1 next-token
probability space. The interaction terms are not displayed for space reasons but are well considered in the
decision trees.}
\label{tab:crossmodel_rf_firstprob_6panel}
\centering
\begin{minipage}{0.48\textwidth}
\centering
\textbf{Qwen-2.5-1.5B-Instruct}\\
\begin{tabular}{p{0.55\linewidth} c c}
\toprule
Model & $R^2_{\mathrm{Adj}}$ & Bal. Acc. \\
\midrule
Top-50 Tokens & 0.273 & 0.607 \\
Top-50 Tokens + Family & 0.387 & 0.566 \\
Top-50 Tokens + Region & 0.331 & 0.621 \\
Top-50 Tokens + Family + Region & 0.413 & 0.575 \\
Top-50 Tokens + Family + Region + Llama Guard & 0.412 & 0.568 \\
\bottomrule
\end{tabular}
\end{minipage}\hfill
\begin{minipage}{0.48\textwidth}
\centering
\textbf{Llama-3.2-1B-Instruct}\\
\begin{tabular}{p{0.55\linewidth} c c}
\toprule
Model & $R^2_{\mathrm{Adj}}$ & Bal. Acc. \\
\midrule
Top-50 Tokens & 0.310 & 0.593 \\
Top-50 Tokens + Family & 0.469 & 0.535 \\
Top-50 Tokens + Region & 0.393 & 0.542 \\
Top-50 Tokens + Family + Region & 0.486 & 0.527 \\
Top-50 Tokens + Family + Region + Llama Guard & 0.486 & 0.526 \\
\bottomrule
\end{tabular}
\end{minipage}
\vspace{0.5em}
\begin{minipage}{0.48\textwidth}
\centering
\textbf{Qwen-2.5-3B-Instruct}\\
\begin{tabular}{p{0.55\linewidth} c c}
\toprule
Model & $R^2_{\mathrm{Adj}}$ & Bal. Acc. \\
\midrule
Top-50 Tokens & 0.155 & 0.584 \\
Top-50 Tokens + Family & 0.291 & 0.581 \\
Top-50 Tokens + Region & 0.213 & 0.702 \\
Top-50 Tokens + Family + Region & 0.311 & 0.576 \\
Top-50 Tokens + Family + Region + Llama Guard & 0.309 & 0.571 \\
\bottomrule
\end{tabular}
\end{minipage}\hfill
\begin{minipage}{0.48\textwidth}
\centering
\textbf{Llama-3.2-3B-Instruct}\\
\begin{tabular}{p{0.55\linewidth} c c}
\toprule
Model & $R^2_{\mathrm{Adj}}$ & Bal. Acc. \\
\midrule
Top-50 Tokens & 0.181 & 0.526 \\
Top-50 Tokens + Family & 0.331 & 0.514 \\
Top-50 Tokens + Region & 0.313 & 0.515 \\
Top-50 Tokens + Family + Region & 0.386 & 0.508 \\
Top-50 Tokens + Family + Region + Llama Guard & 0.385 & 0.508 \\
\bottomrule
\end{tabular}
\end{minipage}
\vspace{0.5em}
\begin{minipage}{0.48\textwidth}
\centering
\textbf{Qwen-2.5-7B-Instruct}\\
\begin{tabular}{p{0.55\linewidth} c c}
\toprule
Model & $R^2_{\mathrm{Adj}}$ & Bal. Acc. \\
\midrule
Top-50 Tokens & 0.158 & 0.628 \\
Top-50 Tokens + Family & 0.273 & 0.696 \\
Top-50 Tokens + Region & 0.202 & 0.724 \\
Top-50 Tokens + Family + Region & 0.289 & 0.679 \\
Top-50 Tokens + Family + Region + Llama Guard & 0.287 & 0.671 \\
\bottomrule
\end{tabular}
\end{minipage}\hfill
\begin{minipage}{0.48\textwidth}
\centering
\textbf{Llama-3.1-8B-Instruct}\\
\begin{tabular}{p{0.55\linewidth} c c}
\toprule
Model & $R^2_{\mathrm{Adj}}$ & Bal. Acc. \\
\midrule
Top-50 Tokens & 0.162 & 0.566 \\
Top-50 Tokens + Family & 0.315 & 0.554 \\
Top-50 Tokens + Region & 0.221 & 0.627 \\
Top-50 Tokens + Family + Region & 0.340 & 0.568 \\
Top-50 Tokens + Family + Region + Llama Guard & 0.338 & 0.578 \\
\bottomrule
\end{tabular}
\end{minipage}

\end{table}

All $5$ regression models suggest only one optimal threshold and so $2$ clusters. However, the associated balanced accuracies are not satisfactory.
We notice the small $R^{2}_{\text{Adj}}$ over all regressions and models leading us to interpret the top-1 next-token probability space as a more complex space than expected.

\newpage
\paragraph{Model's Behavior.} We conduct the GEE logistic regression on all jailbreak prompts to uncover systematic associations between specific variables and the model's behavioral robustness defined by the Llama Guard 4 labels. We test the following $3$-variable model,
\begin{align*}
\text{Safety} \sim \text{Top-1 Probability Cluster} + \text{Paraphrase Family} + \text{Top-6 First Next-Token of Jailbreak Queries}.
\end{align*}
We choose as reference for each three factors the following categories: the low-probability cluster, the Synonyms family and all other first-next-token set (complementary set of the top-6 ones). The dependent variable is the probability of an unsafe answer.
\begin{table}[H]
\caption{GEE logistic regression results. Negative coefficients are associated with a safe label while positive coefficients are associated with an unsafe, compliant answer. P-value* refers to Bonferroni corrected for $10$ comparisons.}
\label{tab:crossmodel_gee_6panel}
\centering
\begin{minipage}{0.48\textwidth}
\centering
\textbf{Qwen-2.5-1.5B-Instruct}\\
\begin{tabular}{l c c}
\toprule
Term & Coefficient & p-value* \\
\midrule
Intercept & -0.606 & <0.001 \\
P Cat 1 & 0.016 & 1.000 \\
Family Leet Speak & -2.744 & <0.001 \\
Family Numbers & -1.751 & <0.001 \\
Family Letter Swap & -0.667 & <0.001 \\
Token [220:Ġ] & -0.132 & 1.000 \\
Token [22555:ĠSure] & 0.512 & <0.001 \\
Token [358:ĠI] & -0.266 & 1.000 \\
Token [508:Ġ[] & 0.064 & 1.000 \\
Token [8082:ĠSur] & -0.245 & 1.000 \\
Token [8835:Ã¢] & -0.131 & 1.000 \\
\bottomrule
\end{tabular}
\end{minipage}
\hfill
\begin{minipage}{0.48\textwidth}
\centering
\textbf{Llama-3.2-1B-Instruct}\\
\begin{tabular}{l c c}
\toprule
Term & Coefficient & p-value* \\
\midrule
Intercept & -3.129 & <0.001 \\
P Cat 1 & -0.011 & 1.000 \\
Family Leet Speak & -0.835 & 0.010 \\
Family Numbers & -1.651 & <0.001 \\
Family Letter Swap & 0.042 & 1.000 \\
Token [11:,] & -1.986 & <0.001 \\
Token [220:Ġ] & -0.148 & 1.000 \\
Token [271:ĊĊ] & -1.058 & <0.001 \\
Token [358:ĠI] & 0.469 & 1.000 \\
Token [40:I] & -0.144 & 1.000 \\
Token [9011:Ã¢] & 1.090 & 0.250 \\
\bottomrule
\end{tabular}
\end{minipage}

\vspace{0.6em}

\begin{minipage}{0.48\textwidth}
\centering
\textbf{Qwen-2.5-3B-Instruct}\\
\begin{tabular}{l c c}
\toprule
Term & Coefficient & p-value* \\
\midrule
Intercept & 0.239 & 0.710 \\
P Cat 1 & -0.025 & 1.000 \\
Family Leet Speak & -1.784 & <0.001 \\
Family Numbers & -1.022 & <0.001 \\
Family Letter Swap & -0.330 & <0.001 \\
Token [198:Ċ] & 0.223 & 1.000 \\
Token [22555:ĠSure] & 0.199 & 0.130 \\
Token [508:Ġ[] & 0.320 & 0.130 \\
Token [8082:ĠSur] & 0.132 & 1.000 \\
Token [82639:Ġ<|] & 0.034 & 1.000 \\
Token [8835:Ã¢] & 0.254 & 1.000 \\
\bottomrule
\end{tabular}
\end{minipage}
\hfill
\begin{minipage}{0.48\textwidth}
\centering
\textbf{Llama-3.2-3B-Instruct}\\
\begin{tabular}{l c c}
\toprule
Term & Coefficient & p-value* \\
\midrule
Intercept & -1.462 & <0.001 \\
P Cat 1 & -0.177 & 1.000 \\
Family Leet Speak & -1.995 & <0.001 \\
Family Numbers & -2.297 & <0.001 \\
Family Letter Swap & -0.306 & 0.050 \\
Token [220:Ġ] & 0.218 & 0.760 \\
Token [271:ĊĊ] & -0.431 & 1.000 \\
Token [4815:ĠĊĊ] & -0.132 & 1.000 \\
Token [510:Ġ[] & -0.043 & 1.000 \\
Token [662:Ġ.] & -0.343 & 1.000 \\
Token [720:ĠĊ] & -0.035 & 1.000 \\
\bottomrule
\end{tabular}
\end{minipage}

\vspace{0.6em}

\begin{minipage}{0.48\textwidth}
\centering
\textbf{Qwen-2.5-7B-Instruct}\\
\begin{tabular}{l c c}
\toprule
Term & Coefficient & p-value* \\
\midrule
Intercept & 0.478 & <0.001 \\
P Cat 1 & 0.092 & 0.280 \\
Family Leet Speak & -1.029 & <0.001 \\
Family Numbers & -0.701 & <0.001 \\
Family Letter Swap & -0.251 & <0.001 \\
Token [198:Ċ] & -0.079 & 1.000 \\
Token [22555:ĠSure] & 0.156 & 1.000 \\
Token [358:ĠI] & -0.200 & 1.000 \\
Token [366:Ġ] & -0.012 & 1.000 \\
Token [508:Ġ[]] & 0.210 & 0.600 \\
Token [8082:ĠSur] & 0.142 & 1.000 \\
\bottomrule
\end{tabular}
\end{minipage}
\hfill
\begin{minipage}{0.48\textwidth}
\centering
\textbf{Llama-3.1-8B-Instruct}\\
\begin{tabular}{l c c}
\toprule
Term & Coefficient & p-value* \\
\midrule
Intercept & -0.724 & <0.001 \\
P Cat 1 & 0.029 & 1.000 \\
Family Leet Speak & 0.013 & 1.000 \\
Family Numbers & -0.666 & <0.001 \\
Family Letter Swap & -0.157 & 1.000 \\
Token [198:Ċ] & -0.514 & 0.010 \\
Token [23371:ĠSure] & -0.150 & 1.000 \\
Token [366:Ġ<] & -1.045 & <0.001 \\
Token [662:Ġ.] & 0.422 & 0.780 \\
Token [720:ĠĊ] & 0.254 & 1.000 \\
Token [9011:Ã¢] & 0.361 & 1.000 \\
\bottomrule
\end{tabular}
\end{minipage}
\end{table}

All three Qwen models and Llama 3.1 occasionally output the first next-token \enquote{Sure} while both Llama 3.2 models are less friendly aligning with their $3.2$ generation safety fine-tuning (\cite{llama32}). There is a general trend for noisy prompts such as Numbers and Leet Speak to be associated with a safe answer, except for Llama 3.1 with respect to the Leet Speak perturbations. This observation may be due to its ability to read complex text. Lastly, the main effect of the top-1 probability cluster does not appear to be significant. Although we hypothesized that the interaction between the top-1 probability and its corresponding token could be informative, this was not feasible to test due to small sample sizes.

\section{AI-Use Disclosure}
Concerning the literature review, we used OpenAI's ChatGPT as a paper searching device according to broad-to-specific instructions. For the analytical tools, we brainstormed with ChatGPT on the SVM analysis, the Random-Forest regression and the GEE logistic regression. 5.3-Codex Medium was requested to write the python scripts according to our instructions. Anthropic's Claude was used as a final revision tool.

\end{document}